\documentclass[a4paper,11pt]{article}
\pdfoutput=1 

\usepackage{jheppub} 
\usepackage{slashed}
\usepackage{mathtools}
\usepackage{amsfonts}
\usepackage{subcaption}
\usepackage{comment}
\usepackage[T1]{fontenc} 

\title{\boldmath Fermi arc in $p$-wave holographic superconductors}

\author{Debabrata Ghorai,}
\author{Taewon Yuk,}
\author{Sang-Jin Sin}
\affiliation{Department of Physics, Hanyang University, Seoul 04763, Korea}

\emailAdd{dghorai123@gmail.com}
\emailAdd{tae1yuk@gmail.com}
\emailAdd{sangjin.sin@gmail.com}

\abstract{ We have investigated the fermionic spectral function in $p$-wave holographic superconductors. We show that the vector model with minimal coupling reveals a $p$-wave spectral function with Fermi arc. This should be contrasted with the previous investigation where $p$-wave arc was demonstrated in the presence of a tensor field. We study the momentum dependent order parameter, the $\omega$-gap in the real part of the conductivity and the fermion spectral function. In addition, we juxtapose the fermionic spectral gap with the order parameter in the holographic set. We demonstrate the impact of coupling constants, temperature and chemical potential on the spectral function.

}

\begin{document} 
	\maketitle
	
	\section{Introduction}	
	\noindent Over the past two decades, the gauge/gravity duality \cite{Maldacena:1997re, Gubser:1998bc, Witten:1998qj} has emerged as a powerful tool for investigating strongly coupled condensed matter systems \cite{HKMS, Seo:2016vks, Oh:2021xbe,Song:2019asj,Oh:2020cym,Oh:2018wfn,Seo:2017oyh,Seo:2018hrc,Seo:2017yux}. This duality connects a strongly coupled boundary field theory with a weakly coupled gravity theory, providing a powerful tool for exploring systems that are difficult to probe otherways. The $U(1)$ symmetry of an Abelian Higgs model in Anti-de Sitter (AdS) spacetime experiences spontaneous symmetry breaking, as described in \cite{Gubser:2008px}. This phenomenon has been utilized in \cite{Hart2008} to uncover fundamental characteristics of high-temperature superconductors, which involve isotropic systems known as $s$-wave holographic superconductors. Moreover, in order to account for $p$-wave superconductivity, one has to introduce non-Abelian gauge fields  \cite{Gubser:2008wv} or a massless vector field \cite{Vegh:2010fc}.  Similarly, a $d$-wave   superconductor  \cite{Romes:2007} should be setup using a rank two tensor field \cite{Benini_2010,Chen:2010mk}.  In our previous study \cite{Ghorai:2022gzx}, we provided an explanation for why we need  vector  field to model $p$-wave superconductivity: only in the setup where we use vector field as superconducting order parameter, the $p$-wave state appears as a ground state. 
Despite a plethora of investigations \cite{Horo2009,Horowitz:2009ij,Horo2011,Ammon:2010pg,Gubser:2010dm,Sin:2009wi,Brihaye:2010mr,Siop2010,Zeng:2010vp,Siopsis:2011vs,zeng2011analytical,Pan:2012jf,Siop2012,gangopadhyay2012analytic,Kim:2013oba,Cai_2014,DeWolfe:2016rxk,ghorai2016higher,Ghorai:2016tvk,Srivastav:2019tkr,Ghorai:2021uby,Donini:2021kne} on holographic superconductors, none have focused on the momentum dependent order parameter, the superconducting gap function.
 Although the fermion spectrum coupled with a vector (or tensor) field setup   allows  Fermi arc in fermionic spectral function\cite{Vegh:2010fc} (\cite{benini2011holographic}), neither the  momentum dependent dispersion relation of the fermion nor the gap function itself have been  demonstrated so far in the literature.  The original meaning of the $p, d$-wave superconductivity means the momentum and angular dependence of the gap function along the   Fermi surface. However,  in the holographic context, previous works used homogeneous components of a vector field as a gap function, leading to some confusion, and therefore, the angular dependence of the gap function has not been clear. \\
			 
    \noindent In this paper, we demonstrate the consistency of the  Fermi arc structure in  the fermionic spectral function  with the  structure of the  radial component of the vector fields, which we identified in  \cite{Ghorai:2022gzx} as the correct order parameter having proper angular dependence.    This is the primary motivation of our work.
     Recently, the fermionic gap for $s$-wave holographic superconductor has been explicitly studied in \cite{Yuk:2022lof}. Previously, the author of ref.\cite{Vegh:2010fc} discussed $p$-wave Fermi arc in the presence of the rank two tensor field in the probe limit, which is not proper to discuss $p$-wave arc structure. 
	\noindent  Our analysis involves two types of interactions with fermion:  the vector interaction and the derivative vector interaction. 
 These two interactions are unique in the sense that they are the simplest interactions  that can exhibit a $p$-wave fermi arc in the fermion spectral function. Our fermion spectral function can  in principle be compared with the  Fermi arc in angle-resolved photoemission spectroscopy (ARPES) data. However, so far, although $p$-wave superconductivity gap is expected from Fe-based superconductors, the data confirming this   is not available yet. This situation should be compared with that of the d-wave case where  many ARPES data are available.\\
 
 	\noindent 
 We will first present the model and its angle-dependent order parameter. We then investigate the AC conductivity  in the fully backreacted system. We will recover the $\omega$-gap in the real part of the conductivity, which is proportional to the value of the order parameter.  
We then discuss the fermionic setup using the flow equation.  We observe that the Fermi arc  in the spectral function is located such that it is consistent with the solution of the gap equation.  
 We also compare the probe limit results with that of the backreacted. Furthermore, our investigation has focused on analyzing the impact of the order parameter value and coupling strength on the $\omega$-gap observed in the fermionic spectral function.
We also examine the effect of temperature and chemical potential on the spectral function, consistent with intuition from bosonic calculations. \\

	\noindent The organization of this paper is as follows. In Section 2, we present a comprehensive discussion of the fundamental framework for the bosonic sector of $p$-wave holographic superconductors.  Section 3 is devoted to a detailed analysis of the fermionic setup for the vector model and the derivation of the flow equation for the bulk Green's function. In Section 4, we examine the fermionic spectral function, providing a comprehensive analysis of our findings. Finally, Section 5 presents a summary of our results.

	\section{Holographic $p$-wave superconductor}
    \noindent The holographic $p$-wave superconductors model can be formulated by introducing a charged vector field that is coupled with the gauge field \cite{cai2015introduction}. This construction is analogous to the non-Abelian gauge field model for holographic superconductors \cite{Sword:2022oyg}. The action can be expressed as follows:
	\begin{eqnarray}
	S_b = \int d^4x \sqrt{-g} \left[\frac{1}{2\kappa^2} \left(R -2 \Lambda\right) + \mathcal{L}_v \right]
	\label{action1}
	\end{eqnarray}
where $\Lambda=-\frac{3}{L^2}$ is the cosmological constant and $L$ is the AdS radius. The matter Lagrangian density for vector field reads
	\begin{eqnarray}
		\mathcal{L}_{v}= -\frac{1}{4}F_{\mu\nu}F^{\mu\nu}-\frac{1}{2}V^{\dagger}_{\mu\nu}V^{\mu\nu}- m^2 V^{\dagger}_{\mu}V^{\mu} + i q_{v} \tilde{\kappa} V_{\mu} V_{\nu}^{\dagger} F^{\mu\nu} 
	\label{pwavev0}
	\end{eqnarray}
	where $F_{\mu\nu}=\partial_{\mu}A_{\nu}-\partial_{\nu}A_{\mu}$ is field strength tensor, $m$ and $q_v$ are the mass and charge of the vector field, respectively. The covariant derivative of the vector field is defined by
	\begin{eqnarray}
	V_{\mu\nu}=\partial_{\mu}V_{\nu}-\partial_{\nu}V_{\mu}-i q_v A_{\mu}V_{\nu}+i q_v A_{\nu}V_{\mu} ~.
	\end{eqnarray}
	From the action (\ref{action1}), the vector field and gauge field equations read
	\begin{eqnarray}
		\label{pwavev1}
		\frac{\partial_{\mu}[\sqrt{-g}F^{\mu\nu}]}{\sqrt{-g}}+ i q_{v} [V^{\dagger}_{\mu} V^{\mu\nu}- V_{\mu}(V^{\mu\nu})^{\dagger}] + \frac{ i q_{v} \tilde{\kappa}}{\sqrt{-g}} \partial_{\mu}\left[ \sqrt{-g} \left( V^{\mu\dagger}V^{\nu} - V^{\mu} V^{\nu\dagger} \right) \right] &=& 0 ~,~~~ \\
		\frac{1}{\sqrt{-g}}\partial_{\mu}[\sqrt{-g}V^{\mu\nu}]- [m^2 V^{\nu}+i q_{v} A_{\mu} V^{\mu\nu}]+ i q_{v} \tilde{\kappa} V_{\mu} F^{\mu\nu} &=& 0 ~.
\label{pwavev2}
	\end{eqnarray}
The Einstein field equation reads
	\begin{eqnarray}
		&& R_{\mu\nu} -\frac{1}{2} g_{\mu\nu} R + \Lambda g_{\mu\nu} = 2\kappa^2 T_{\mu\nu} \nonumber \\
		T_{\mu\nu} &=& \frac{1}{2} g_{\mu\nu} \mathcal{L}_p^{V} + \frac{1}{2} F_{\mu\lambda} F^{\lambda}_{\nu} +\frac{1}{2} \left[ V^{\dagger}_{\mu\beta} V^{\beta}_{\nu} + m^2 V_{\mu}^{\dagger}V_{\nu} -i q_{v} \tilde{\kappa} (V_{\mu}V^{\dagger}_{\beta} - V^{\dagger}_{\mu} V_{\beta})F_{\nu}^{\beta} + \mu \leftrightarrow \nu \right] ~~~
	\end{eqnarray}

    \noindent In order to account for the effects of backreaction and spacetime anisotropy, we adopt the use of an anisotropic metric along the $xy$ plane, which corresponds to the coordinate system of the boundary field theory and is given by 
    \begin{eqnarray}
    	ds^2 = \frac{1}{z^2} \left[- f(z) e^{-\chi (z)} dt^2 + \frac{dz^2}{f(z)} + h^2(z) dx^2 + \frac{1}{h^2(z)} dy^2 \right] ~.
    \end{eqnarray}
    {The above   metric is asymptotically AdS since  $g_{\mu\nu} \rightarrow \frac{\eta_{\mu\nu}}{z^2}$ for $z\rightarrow 0$, where $\eta_{\mu\nu}$ is the Minkowski metric.}
    The Hawking temperature reads in $z$ coordinate 
    \begin{eqnarray}
    	T_H = -\frac{f'(z_h) e^{-\frac{\chi(z_h)}{2}}}{4\pi} ~,
    	\label{hawkingt1}
    \end{eqnarray}
where $z_h$ is the horizon radius. To break the rotational symmetry of the vector field in $xy$ plane, one can consider a complex vector field $V_{\mu}dx^{\mu}=V_i (z,x,y)dx^{i}$ along the $i$-direction (where $i$ can be either $x$ or $y$). In a previous study \cite{Ghorai:2022gzx}, we have demonstrated that the ground state of the vector field model corresponds to a $p$-wave state and the corrected order parameter is $V_u$ (where $u$ is the radial coordinate in the $xy$-plane) instead of $V_x$ or $V_y$ since $V_u$ depends on the angle.  Furthermore, we showed that the solution for $V_i(z,x,y)dx^{i}$ reduces to $V_{i}(z)dx^{i}$ only for the ground state. Given our focus on the ground state configuration of the vector field, we adopt the vector ansatz $V= V_{y}(z) dy$ and the gauge field ansatz $A=A_t(z)dt$. Therefore, we can express the vector field as:
	\begin{eqnarray}
		V= V_{y}(z) dy = V_u du + V_{\theta} d\theta 
	\end{eqnarray}
	where $V_u = V_{y}(z) \sin\theta $ and $V_{\theta} = V_{y}(z) u \cos\theta $. Since $V_{\theta}$ has coordinate singularity at $u=0$, the $V_u$ is the corrected order parameter for $p$-wave holographic superconductors. \\

	\noindent We now try to calculate all fields configuration with backreaction. Using the above vector field and gauge field ansatz, all fields equations reads
	\begin{eqnarray}
		\label{pbeom1}
		h''+ \left[\frac{f'}{f}-\frac{\chi'}{2}-\frac{2 }{z}-\frac{h'}{h}\right]h'+
		\kappa^2 h^3 z^2\left[\left(\frac{q_v^2 e^{\chi} A_t^2}{f^2} -\frac{m^2}{z^2f}\right)V_y^2  -  V_y^{\prime 2}\right] &=& 0  \\
		\label{pbeom1a}
		\chi'-\frac{f'}{f}-\frac{z h^{\prime 2}}{2 h^2}-\frac{3}{L^2 z f}+\frac{3}{z} +\kappa^2 h^2 z^3 \left[ \frac{e^{\chi} A_t^{\prime 2}}{2 f h^2}- \left(\frac{q_v^2 e^{\chi} A_t^2}{f^2} -\frac{m^2}{z^2f}\right)V_y^2  -  V_y^{\prime 2} \right] &=& 0  \\
		\label{pbeom1b}
		f'-\left[\frac{z  h^{\prime 2}}{h^2}+\frac{3 }{z}\right]f +\frac{3}{L^2 z} - \kappa^2 h^2 z^3 \left[\frac{e^{\chi} A_t^{\prime 2}}{2 h^2}+ \left(\frac{q_v^2 e^{\chi} A_t^2}{f} +\frac{m^2}{z^2}\right)V_y^2  + f V_y^{\prime 2}\right] &=& 0  \\
		\label{pbeom1c}
		A_t^{\prime\prime}+\frac{\chi^{\prime}}{2}A_t^{\prime}-\frac{2 q_v^2 h^2 V_y^2}{f} A_t &=& 0  \\
		\label{pbeom1d}
		V_y^{\prime\prime}+ \left[\frac{f'}{f}+\frac{2h'}{h}-\frac{\chi^{\prime}}{2} \right]V_y^{\prime}+\left[\frac{q_v^2 e^{\chi} A_t^2}{f^2}-\frac{m^2}{z^2 f}\right]V_y &=& 0 ~~~
	\end{eqnarray}
	where prime denotes the derivative with respect to $z$ coordinate. Using eq.(\ref{pbeom1b}), we find
	\begin{eqnarray}
		f^{\prime} (z_h)= \kappa^2\left(\frac{1}{2} e^{\chi(z_h)} z_h^3 A_t^{\prime}(z_h)^2 + m^2 z_h h(z_h)^2 V_y(z_h)^2 \right)- \frac{3}{L^2 z_h} ~,
	\end{eqnarray}
    which leads to determine the Hawking temperature (\ref{hawkingt1}) as follows
    \begin{eqnarray}
    	T_H = \frac{3}{4\pi L^2 z_h} \left[1-\frac{\kappa^2 L^2}{3}\left(\frac{e^{\chi(z_h)}}{2}  z_h^4 A_t^{\prime}(z_h)^2 + m^2 z_h^2 h(z_h)^2 V_y(z_h)^2 \right) \right]e^{-\frac{\chi(z_h)}{2}} ~.
    	\label{hawkingtemp2}
    \end{eqnarray}
    \noindent The spacetime is asymptotically AdS which imposes the boundary condition on $h(z=0)=1$ and $\chi(z=0)=0$. Therefore the gauge field and vector field equations at the boundary become
    \begin{eqnarray}
    A^{\prime\prime}(z) =0 ~~~~~\text{and}~~~~~~~~ V_y^{\prime\prime}(z) -\frac{m^2}{z^2} V_y(z)=0 ~.
    \end{eqnarray} 
    which gives the asymptotic behaviour of the gauge field and vector field in terms of quantities of boundary theory in following way: 
    \begin{eqnarray}
    	A_t(z) = \mu -\rho z ~~~\text{and} ~~~ V_y(z) = C_s z^{\delta_{-}-1} + C_c z^{\delta_{+}-1}
    \end{eqnarray}
    where $\mu, \rho, C_s, C_c $ are the chemical potential, charge density, source term and expection value of angle independent condensation of the boundary theory respectively. The scaling dimension $\delta_{\pm}=\frac{1}{2}\left[3\pm \sqrt{1+4 m^2} \right]$ is determined from the mass $m$ of the vector field. The explicit angle dependent order parameter for $p$-wave holographic superconductor is 
        \begin{eqnarray}
     V_u=V_y(z) \sin\theta=C_s \sin\theta z^{\delta_{-}-1} + \langle\mathcal{O}\rangle z^{\delta_{+}-1} ~,
    \end{eqnarray}
    where $\langle\mathcal{O}\rangle= C_c \sin\theta$ is the angle dependent condensation operator value. 
 We set $C_s=0$ since, in the absence of a source, condensation is expected to occur due to $U(1)$ symmetry breaking in the holographic set. 
    In order to determine the angle dependent condensation operator value, we now need to calculate the $C_c$ value in the full backreacted system. This gives us momentum dependent gap function in Fourier space $(k_x, k_y)$ with help of  the following identification:
    \begin{eqnarray}
    	 \Delta_k = FT[\langle\mathcal{O}\rangle]= \frac{1}{2\pi a^2}\int_{0}^{a} \int_{0}^{2\pi} \langle\mathcal{O}\rangle  e^{-i (k_x u\cos\theta + k_y u\sin\theta)} udud\theta  ~,
    	 \label{2dfourier}
    \end{eqnarray} 
     where $a$ is the sample size and $FT[...]$ is the two dimensional Fourier transformation. 
  
	\subsection{The critical temperature and momentum dependent gap structure}
\begin{figure}[h!]
		\centering
		\begin{subfigure}[b]{0.32\textwidth}
			\centering
			\includegraphics[scale=0.23]{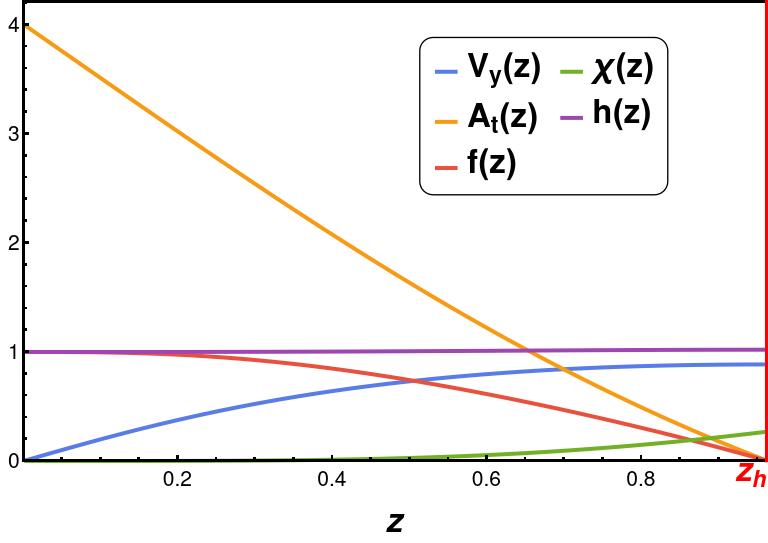}
			\caption{For $T=0.03525 \mu $}
		\end{subfigure}
		\hfil
		\begin{subfigure}[b]{0.32\textwidth}
			\centering
			\includegraphics[scale=0.23]{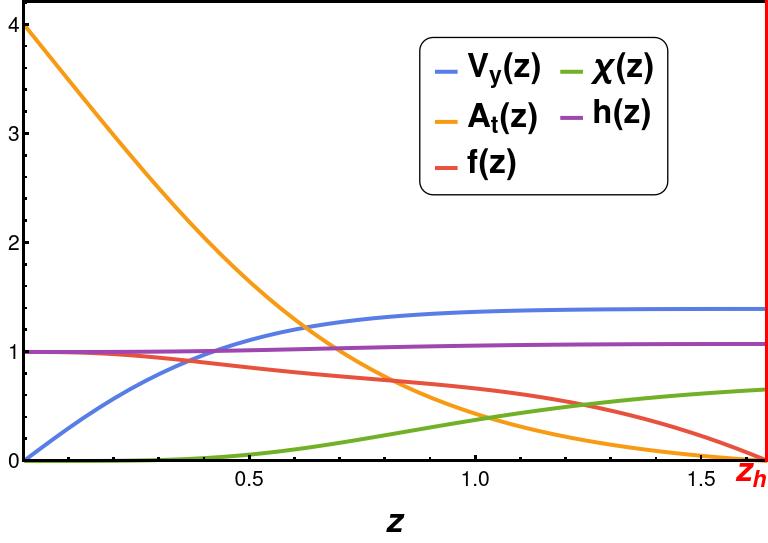}
			\caption{For $T=0.025 \mu$}
		\end{subfigure}
		\hfil
		\begin{subfigure}[b]{0.32\textwidth}
			\centering
			\includegraphics[scale=0.23]{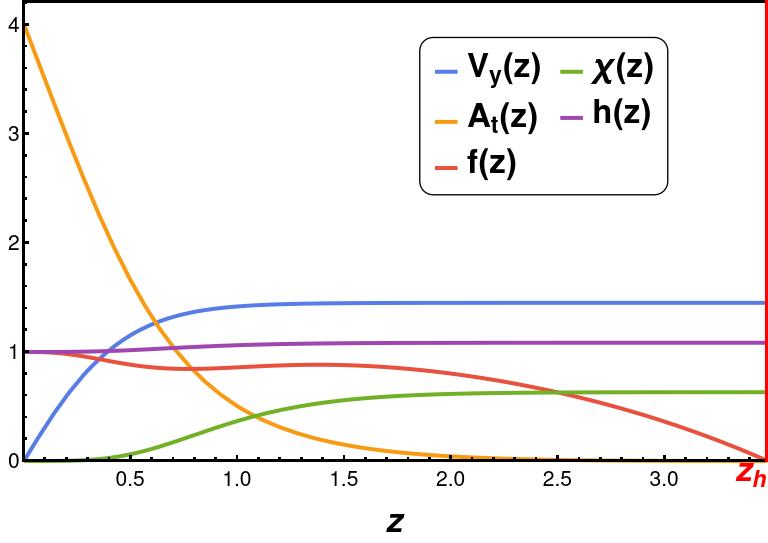}
			\caption{For $T=0.0125 \mu$}
		\end{subfigure}
		\caption{Backreacted profiles at three different temperatures below $T_c=0.03748\mu$.}
		\label{fig1b}
	\end{figure}

\begin{figure}[h!]
		\centering
		\begin{subfigure}[b]{0.49\textwidth}
		\centering
		\includegraphics[scale=0.39]{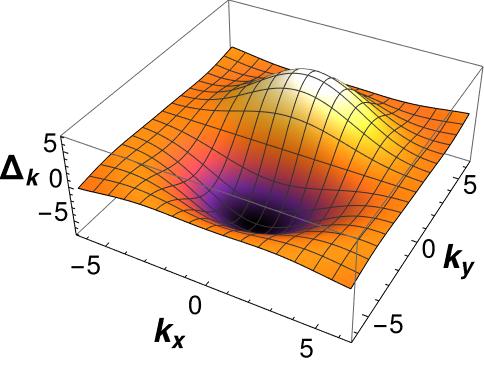}
		\caption{$\Delta_k$ at fixed $T=0.446 T_c$}
		\end{subfigure}
	\begin{subfigure}[b]{0.49\textwidth}
		\centering
		\includegraphics[scale=0.25]{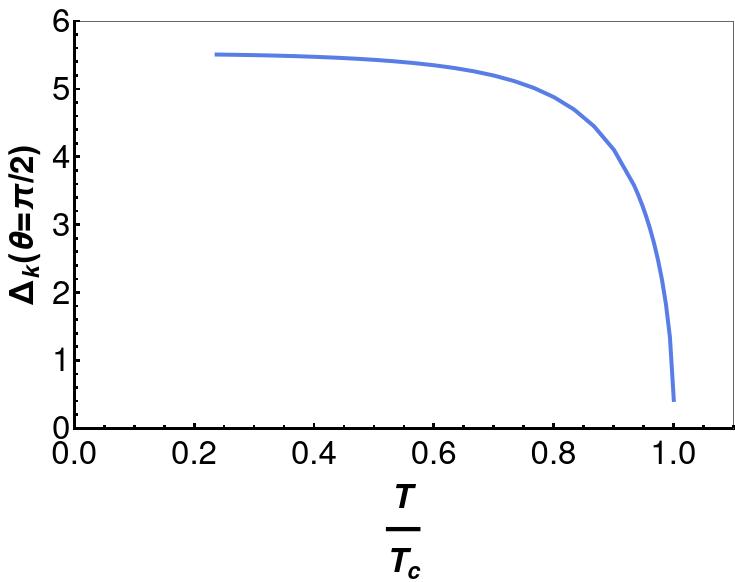}
		\caption{Maximum value of $\Delta_k$ vs $\frac{T}{T_c}$}
		\end{subfigure}
	\caption{Momentum dependent order parameter $\Delta_k=FT [ \langle \mathcal{O}\rangle ]$ for $p$-wave HSC. (a) Three dimensional plot of $\Delta_k$ for fixed temperature $T=0.446 T_c$. (b) Maximum value of $\Delta_k$ vs $\frac{T}{T_c}$ plot. The corresponding $(k_x, k_y)$ value for maximum $\Delta_k$ is $(0, 2.45)$. }
		\label{figgap}
	\end{figure}

     In this subsection, we will numerically solve all field equation by employing Shooting method. To solve all the coupled equations, we need boundary conditions of all field which are
     \begin{align}
     	A_{t}(z_h)=0, ~~~~~f(z_h)=0,  ~~~~~h(0)=1, ~~~~~\chi(0)=0, ~~~~~C_s=0 ~.
     \end{align} 
     Therefore $(T, \mu)$ are the controlling parameters to solve all the field equations.
     The near horizon behaviour of the fields can be obtained from the Taylor series expansion of the fields in following way: 
	\begin{eqnarray}
		\label{hexpansion}
		(V_y(z), A_t(z), f(z), \chi(z), h(z)) \approx \sum_{i=0}^{5} (V_{yi}, A_{ti}, f_{i}, \chi_{i}, h_{i})\left(1-\frac{z}{z_h}\right)^{i} ~~~.
	\end{eqnarray}
    By plugging in the above expansion into the field equations given by (\ref{pbeom1}), we can establish relations between the aforementioned coefficients $(V_{yi}, A_{ti}, f_{i}, \chi_{i}, h_{i})$ in terms of the horizon data $(V_{y0}, A_{t1}, z_h, \chi_{0}, h_{0})$. By imposing the appropriate boundary conditions and solving the equations of motion for the fields, the horizon data can be determined for a desired $(T,\mu)=(T_0, \mu_0)$, enabling the use of the Shooting method. Utilizing these field equations (\ref{pbeom1}-\ref{pbeom1d}) along with the solution of the horizon data yields all field configurations for a desired ratio $\frac{T}{\mu}$.
    Setting $\kappa=1/2, q_v=1, m^2=0, L=1$, and using eq.(\ref{hawkingtemp2}), we have found $T_c=0.03748 \mu$ for the scaling dimension $\delta=\delta_{+}=2$, which matches exactly with previous findings in \cite{Sword:2022oyg}. For $T=0.01\mu$, the condensation value is $C_c \approx 5.38727\mu$. When $T$ is much smaller than $T_c$, all field configurations become highly curved, as shown in Figure \ref{fig1b}.
	 \\
    Using this field solution and eq.(\ref{2dfourier}), we can now plot the momentum dependent gap function at $T=0.446 T_c$ in Figure \ref{figgap}(a), which is along $k_y$ direction. 
     {The maximum value of the gap function is at $(k_x, k_y)=(0, 2.45)$ for any temperature below $T_c$. This  order parameter for fixed momentum decreases as the temperature increases, as shown in Figure \ref{figgap}(b).}
    In a later section, we will examine the fermionic spectral function, which will show the fermionic gap along the $k_y$ direction. This analysis will support the notion that the corrected order parameter in $p$-wave holographic superconductors should be $V_u$.
	 
	 \subsection{Conductivity in the presence of backreacted fields}
\begin{figure}[h!]
		\centering
		\begin{subfigure}[b]{0.32\textwidth}
			\centering
			\includegraphics[scale=0.24]{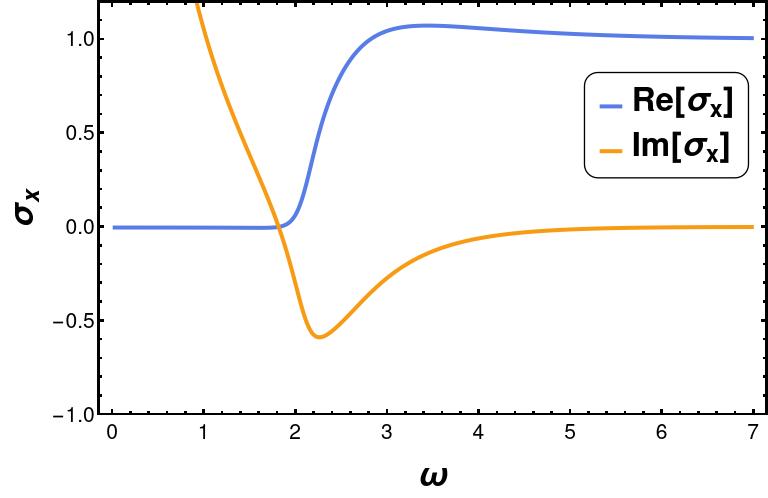}
			\caption{$\sigma_{x}$ at $T=0.267T_c$}
		\end{subfigure}
		\hfil
		\begin{subfigure}[b]{0.32\textwidth}
			\centering
			\includegraphics[scale=0.24]{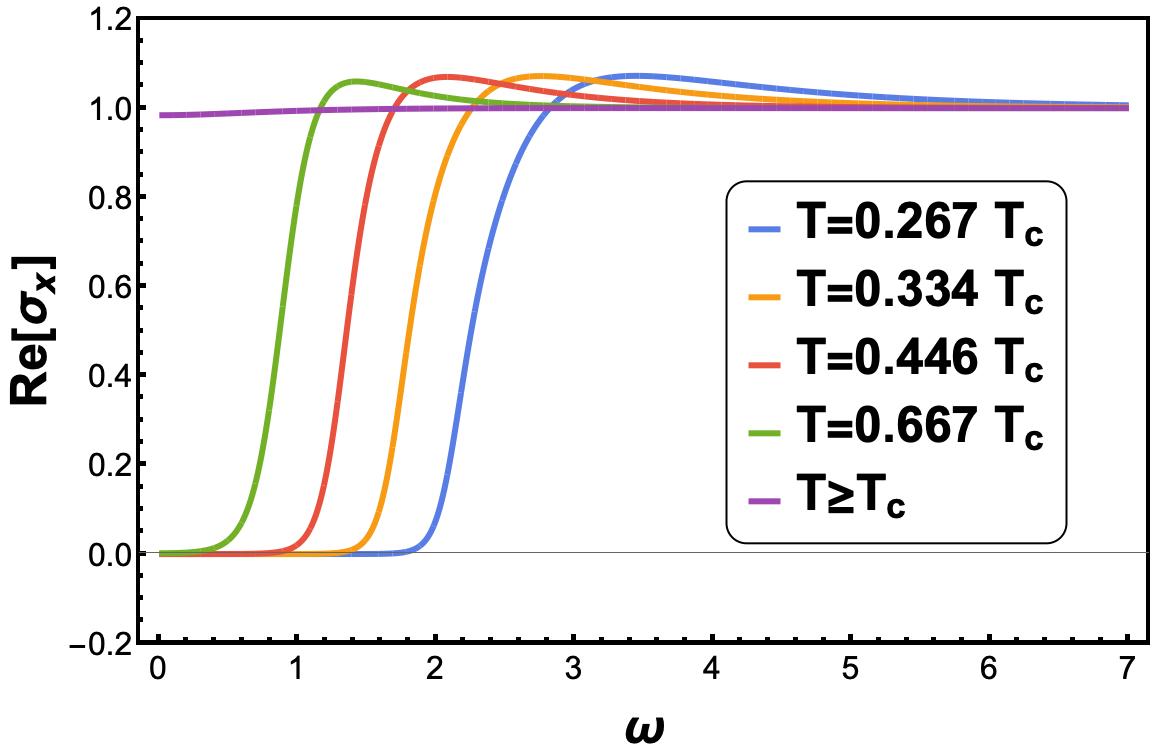}
			\caption{$Re[\sigma_{x}]$ at different $T$}
		\end{subfigure}
		\hfil
		\begin{subfigure}[b]{0.32\textwidth}
			\centering
			\includegraphics[scale=0.237]{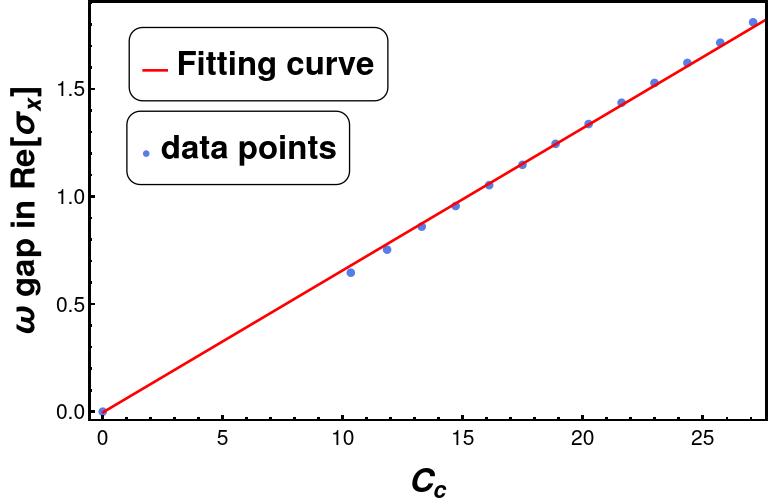}
			\caption{Gap in $Re[\sigma_x]$ vs  $C_c$}
		\end{subfigure}
		\caption{Conductivity $\sigma_{x}$ for the backreacted vector field. (a) Plot of $Re[\sigma_{x}]$ and $Im[\sigma_{x}]$ at fixed temperature $T=0.267T_c$. $Im[\sigma_x]$ has a pole at $\omega=0$. (b) Plot of $Re[\sigma_{x}]$ vs $\omega$ at different temperatures. (c) The gap in the $Re[\sigma_{x}]$ is $0.066 C_c$. }
		\label{figcond}
	\end{figure}

	 The fluctuation of the gauge field along $x$ direction gives the current along $x$ direction, which can be analysed in this holographic set with gauge field ansatz $A_{\mu}=(0, A_x(z) e^{- i \omega t},0,0)$. From eq.(\ref{pwavev2}), we obtain the gauge field equation
	\begin{eqnarray}
		A_x''+ \left[ \frac{ f'}{f}-\frac{\chi'}{2}  -\frac{2 h'}{h}\right]A_x'+ \left[\frac{\omega ^2  e^{\chi}}{f^2}-\frac{2 h^2 V_y^2}{f}\right]A_x =0 ~~.
		\label{cond1}
	\end{eqnarray}
    To solve this equation numerically, we need to find the horizon behaviour of $A_x(z)$ field. The field at near horizon  can be expressed as follows:
    \begin{eqnarray}
    	A_x \approx \left(1-\frac{z}{z_h}\right)^{\omega a_{x0}} \left[1+ a_{x1}\left(1-\frac{z}{z_h}\right)+...+a_{x4}\left(1-\frac{z}{z_h}\right)^4\right] ~~~.
    \end{eqnarray}
    Using the horizon expansion of all fields (\ref{hexpansion}), we can solve $a_{xi}, (i=0,1,2,3,4)$ in terms of all input horizon data $(V_{y0}, A_{t1}, z_h, \chi_{0}, h_{0})$ for fixed values of $\frac{T}{\mu}$ ratio. We can now solve the gauge field equation with the horizon data. From gauge/gravity duality, we can derive the expression of AC conductivity \cite{Hartnoll:2008kx} in following form
    \begin{eqnarray}
    	\sigma_{x} (\omega) =\frac{i}{\omega}\frac{A^{\prime}_x(z)}{A_x(z)}|_{z=0} ~~.
    \end{eqnarray}
    Solving the gauge field eq.(\ref{cond1}), we have computed the AC conductivity for the vector field model, which is shown in Figure \ref{figcond}. From the imaginary part of the conductivity plot, we can conclude the infinite DC conductivity $\sigma_{x}(\omega=0$) using the Kramers$-$Kronig relation since the imaginary part of the conductivity has a pole at $\omega=0$. We also observe a clear frequency gap ($\omega$-gap) in the real part of the AC conductivity, one of the main indicators of superconductor theory. From numerical analysis, we find the $\omega$-gap $ \approx 0.066 C_c$, shown in Figure \ref{figcond}(c).
	 \\
	
\section{Fermion in the $p$-wave superconductor}
We now investigate the fermionic spectral function along with this fully backreacted bosonic configuration. The action for the fermion part is given by
	\begin{eqnarray}
		S_{\psi} &=& \int d^4x \sqrt{-g} \left[i \bar{\psi} (\Gamma^{\mu}D_{\mu}- m_{f})\psi -i \bar{\psi}_c (\Gamma^{\mu}D^{*}_{\mu}- m_{f})\psi_c +\mathcal{L}_{int} \right]\nonumber \\
		\mathcal{L}_{int} &=&
				g_v \bar{\psi} V_{\mu}\Gamma^{\mu} \psi_c - g_{dv} \bar{\psi} V_{\mu\nu}\Gamma^{\mu\nu} \psi_c + h.c. ~,
	\label{fermieq1}
	\end{eqnarray}
	where $g_v, g_{dv}$ are coupling constants, $m_f$ is the mass of the fermion, the covariant derivative of spinor is $D_{\mu}=\partial_{\mu} + \frac{1}{4} \omega_{\mu\bar{\alpha\beta}}\Gamma^{\bar{\alpha\beta}}- i q A_{\mu}$ and $\psi_{ c}=\psi^{*}$ is the complex charge conjugate field of the fermion which is treated as an independent field in all computations in this set-up.  {We have considered two   minimal interaction terms: the vector interaction $g_v \bar{\psi} V_{\mu}\Gamma^{\mu} \psi_c$ and the derivative vector interaction $g_{dv} \bar{\psi} V_{\mu\nu}\Gamma^{\mu\nu} \psi_c$. The driving force behind exploring these two interactions lies in their minimalistic nature as the two feasible vector field interactions with fermions capable of generating a $p$-wave fermionic spectral function. } We choose the following bulk gamma matrices:
	\begin{eqnarray}
		\Gamma^{\underline{t}} = \sigma_1 \otimes i\sigma_2, ~~ \Gamma^{\underline{x}} = \sigma_1 \otimes \sigma_1, ~~ \Gamma^{\underline{y}} = \sigma_1 \otimes \sigma_3, ~~
		\Gamma^{\underline{z}} = \sigma_3 \otimes \sigma_0 ~,
	\end{eqnarray} 
	where underline indices represent tangent space indices. 
	The Dirac equation reads
	\begin{eqnarray}
		(\Gamma^{\mu}D_{\mu}- m_{f})\psi - i g_v  V_{\mu}\Gamma^{\mu} \psi_c + i g_{dv} V_{\mu\nu}\Gamma^{\mu\nu} \psi_c   = 0 ~~~.
	\end{eqnarray}
    For simplicity, we consider the fermionic field in following ways:
	\begin{eqnarray}
		\psi (t, x, y, z) = \frac{1}{(-gg^{zz})^{1/4}} e^{-i \omega t + i k_x x + i k_y y } \Psi(z) ~~~.
		\label{eq26}
	\end{eqnarray}
	Substituting the above spinor in the Dirac equations, we obtain
	\begin{eqnarray}
		\left[\Gamma^{\underline{z}}\partial_z - i \left( \sqrt{\frac{g^{tt}}{g^{zz}}}(\omega+ qA_t)\Gamma^{\underline{t}} - \sqrt{\frac{g^{xx}}{g^{zz}}} k_x \Gamma^{\underline{x}} -  \sqrt{\frac{g^{yy}}{g^{zz}}} k_y \Gamma^{\underline{y}}\right) -\frac{m_f}{\sqrt{g^{zz}}} \right] \Psi (z) \nonumber\\ 
		 - \frac{i}{\sqrt{g^{zz}}}\mathcal{I}_{int} \Psi_c (z) =0   
		\label{eq27a1}
	\end{eqnarray} 
	where $\mathcal{I}_{int}= \left(g_v V_{\mu}\Gamma^{\mu}- g_{dv} V_{\mu\nu}\Gamma^{\mu\nu}\right)$. The field equation for conjugate fermion is 
	\begin{eqnarray}
		\left[\Gamma^{\underline{z}}\partial_z + i \left( \sqrt{\frac{g^{tt}}{g^{zz}}}(\omega+ qA_t)\Gamma^{\underline{t}} - \sqrt{\frac{g^{xx}}{g^{zz}}} k_x \Gamma^{\underline{x}} -  \sqrt{\frac{g^{yy}}{g^{zz}}} k_y \Gamma^{\underline{y}}\right) -\frac{m_f}{\sqrt{g^{zz}}} \right] \Psi_c (z) \nonumber \\+ \frac{i}{\sqrt{g^{zz}}}\mathcal{I}_{int} \Psi (z) =0~.  		\label{eq27b1}
	\end{eqnarray} 
	We can write the $4$-components spinor as 
	\begin{eqnarray}
		\Psi(z)= \begin{pmatrix}
			\Psi_{+}(z) \\
			\Psi_{-}(z)
		\end{pmatrix}, ~~~~~~\text{where}~~ \Psi_{\pm}= \begin{pmatrix}
			\Psi_{\pm 1} \\
			\Psi_{\pm 2}
		\end{pmatrix}
	\end{eqnarray}
	which gives us the Dirac equation in following way
	\begin{eqnarray}
		\left[\partial_z \mp \frac{m_f}{\sqrt{g^{zz}}}\right] \Psi_{\pm} =\pm i \left[K_{a}\gamma^{a}\Psi_{\mp} + \frac{1}{\sqrt{g^{zz}}} \mathcal{I}_{int} \Psi_{c\pm}\right]
		\label{eq39}
	\end{eqnarray}
	where $K_{a}= \left(\sqrt{\frac{g^{tt}}{g^{zz}}} (\omega +q A_t), -\sqrt{\frac{g^{xx}}{g^{zz}}}k_x, -\sqrt{\frac{g^{yy}}{g^{zz}}}k_y\right)$ and $\gamma^{a}=\left(i\sigma_2, \sigma_1, \sigma_3\right) $. Similarly, we can rewrite equation of motion for the conjugate fermion. For the asymptotic limit ($z\rightarrow 0$), we can consider all metric components $g^{\mu\nu}\rightarrow z^2\eta^{\mu\nu}$.
	Therefore, the asymptotic behaviour of the spinor reads
	\begin{eqnarray}
		\Psi_{+} (z) &=& \mathbf{A} z^{m_f} + \mathbf{B} z^{1-m_f},  ~~~~~ \Psi_{-}(z) = \mathbf{D} z^{-m_f} + \mathbf{C} z^{1+m_f}~, \\
		\Psi_{c+} (z) &=& \tilde{\mathbf{A}}^{*} z^{m_f} + \tilde{\mathbf{B}}^{*} z^{1-m_f},  ~~~~~ \Psi_{c-}(z) = \tilde{\mathbf{D}}^{*} z^{-m_f} + \tilde{\mathbf{C}}^{*} z^{1+m_f} ~,
	\end{eqnarray}
	where $\mathbf{A}, \mathbf{B}, \mathbf{C}, \mathbf{D}$ are two component spinors, which are determined by solving the full bulk Dirac equations. For $|m_f|<\frac{1}{2}$, the leading term makes the boundary spinors solution as \cite{Liu:2009dm,Faulkner:2009wj}
	\begin{eqnarray}
		\Psi (z) \approx \begin{pmatrix}
			\mathbf{A} z^{m_f} \\
			\mathbf{D} z^{-m_f} 
		\end{pmatrix},  ~~~~~\Psi_c (z) \approx \begin{pmatrix}
			\tilde{\mathbf{A}}^{*} z^{m_f} \\
			\tilde{\mathbf{D}}^{*} z^{-m_f} 
		\end{pmatrix} ~.
		\label{eq211a}
	\end{eqnarray}
	We have found that the leading order of the asymptotic behaviour of fields is always $z^{\pm m_f}$ for $|m_f|<\frac{1}{2}$ which is independent of these interactions. The two component spinor contains the effect of interactions.

	\subsection{The source and the response of the fermionic field}
	To know the source and response, we need to identify the boundary term of the bulk fermion field. From the variation of the bulk action, we obtain
	\begin{eqnarray}
		\delta S_{bulk} = EoM + i \int d^3x \sqrt{-h} [\bar{\psi} \Gamma^{\underline{z}} \delta\psi -\delta\bar{\psi} \Gamma^{\underline{z}} \psi -\bar{\psi_c} \Gamma^{\underline{z}} \delta\psi_c + \delta\bar{\psi_c} \Gamma^{\underline{z}} \psi_c]   ~.
	\end{eqnarray}
For standard quantization \cite{Iqbal:2009fd}, the boundary action is given by
\begin{eqnarray}
	S_{bdy}= i \int d^3x  \sqrt{-h} (\bar{\psi}\psi+ \bar{\psi}_c \psi_{ c}) ~.
\end{eqnarray}
	The variation of the boundary action is 
	\begin{eqnarray}
		\delta S_{bdy} = i \int d^3x \sqrt{-h} [\delta\bar{\psi} \psi + \bar{\psi}\delta\psi +\delta\bar{\psi_c} \psi_c + \bar{\psi_c}\delta\psi_c] ~.
	\end{eqnarray}
	Using eq.(\ref{eq26}), we obtain the total bulk action variation
	\begin{eqnarray}
		\delta S = i \int d^3x [\bar{\Psi} \Gamma^{\underline{z}} \delta\Psi -\delta\bar{\Psi} \Gamma^{\underline{z}} \Psi -\bar{\Psi}_c \Gamma^{\underline{z}} \delta\Psi_c + \delta\bar{\Psi}_c \Gamma^{\underline{z}} \Psi_c + \delta\bar{\Psi} \Psi + \bar{\Psi}\delta\Psi +\delta\bar{\Psi}_c \Psi_c + \bar{\Psi}_c\delta\Psi_c ] ~~~~~~
		\label{eq211}
	\end{eqnarray}
	where $S=S_{bulk}+ S_{bdy}$. Using two-component spinor structure, the above eq.(\ref{eq211}) becomes
	\begin{eqnarray}
		\delta S &=& 2i \int d^3x  \left[ \Psi_{-}^{\dagger} (i\sigma_2) \delta\Psi_{+} + \delta\Psi_{+}^{\dagger} (i\sigma_2) \Psi_{-} + \Psi_{c+}^{\dagger} (i\sigma_2) \delta\Psi_{c-} + \delta\Psi_{-c}^{\dagger} (i\sigma_2) \Psi_{c+}  \right] .~~~
	\end{eqnarray}
From the above expression, we can say that we need to choose $\xi^{(S)}= \begin{pmatrix} 
		\Psi_{+} \\
		\Psi_{c-}
	\end{pmatrix}$ to be the source whose values are fixed at the boundary. This will make the variation of the total action zero. The remaining components of the spinors can identify as the response or condensate which can be derived from the boundary action
	\begin{eqnarray}
		S_{bdy}= \int d^3x  \left[ \Psi_{-}^{\dagger}(-\sigma_2)\Psi_{+} +\Psi_{c+}^{\dagger}(-\sigma_2)\Psi_{c-} + h.c. \right] .
		\label{eq217}
	\end{eqnarray}
	From the above expression, the condensation is  
	\begin{eqnarray}
		\xi^{(C)}= \begin{pmatrix} 
			\Psi_{-} \\
			\Psi_{c+}
		\end{pmatrix},  ~~~~~~~\text{since}~~ \xi^{(S)}= \begin{pmatrix} 
			\Psi_{+} \\
			\Psi_{c-}
		\end{pmatrix}.
		\label{eq319}
	\end{eqnarray}
	Then the boundary action can be written as
	\begin{eqnarray}
		S_{bdy} = \int d^3x  \left[ \xi^{(C)\dagger} \tilde{\Gamma}\xi^{(S)}+ \xi^{(S)\dagger} \tilde{\Gamma}\xi^{(C)} \right]
		\label{eq218}
	\end{eqnarray}
	where the boundary gamma matrix $\tilde{\Gamma}= \sigma_0 \otimes (-\sigma_2) $.  {This $\xi^{(C)}$ and $\xi^{(S)}$ are the Nambu-Gorkov spinors that are used to describe systems with   particle-hole symmetry. This spinor is the valuable basis for the representation of BCS mean-field Hamiltonian \cite{coleman_2015}.}

	\subsection{Green's function from flow equation}
	Rearranging all components of eq.(s)(\ref{eq27a1},\ref{eq27b1}), we can recast the Dirac equations in following structure
	\begin{eqnarray}
		\label{eq215}
		\partial_z \xi^{(S)} + \mathbb{M}_1 \xi^{(S)}  + \mathbb{M}_2 \xi^{(C)} &=& 0 ~, \\
		\partial_z \xi^{(C)} + \mathbb{M}_3 \xi^{(C)}  + \mathbb{M}_4 \xi^{(S)} &=& 0 
		\label{eq216}
	\end{eqnarray}
	where $4\times4$-matrix $\mathbb{M}_{i}, ~i=1,2,3,4$ are determined from (\ref{eq27a1},\ref{eq27b1}). We have calculated those $\mathbb{M}_{i}$ which are 
		\begin{eqnarray}
		\mathbb{M}_1= \begin{pmatrix}
			\mathbb{N}_1 & \mathbb{P}_1 (g_{v}) \\
			\mathbb{P}_1(g_{v}) & - \mathbb{N}_1
		\end{pmatrix}, 		\mathbb{M}_2= \begin{pmatrix}
		\mathbb{N}_2(q) & -\mathbb{P}_2 \\
		\mathbb{P}_2 & \mathbb{N}_2 (-q)
	\end{pmatrix}, 	\mathbb{M}_3= \begin{pmatrix}
	-\mathbb{N}_1 & \mathbb{P}_1 (-g_{v}) \\
	\mathbb{P}_1(-g_{v}) & \mathbb{N}_1
\end{pmatrix}, \mathbb{M}_4= -\mathbb{M}_2 ~~~~~~~
	\end{eqnarray}
where
	\begin{eqnarray}
 \mathbb{N}_1 = - \frac{m_f}{z \sqrt{f}}\boldsymbol{1}_{2\times 2}, ~~~~\mathbb{N}_2 (q) = \frac{i}{\sqrt{f}} \begin{pmatrix}
			k_y h & -\frac{e^{\chi/2} (\omega+ q A_t)}{\sqrt{f}}+\frac{k_x}{h}\\
			\frac{e^{\chi/2} (\omega+ q A_t)}{\sqrt{f}}+\frac{k_x}{h} & -k_y h
		\end{pmatrix} ~, \nonumber \\
	\mathbb{P}_1(g_{v})=h\begin{pmatrix}
		2 i g_{dv} z  V_y^{\prime}-\frac{i g_v  V_y}{\sqrt{f}} & 0 \\
		0 &  -2 i g_{dv} z  V_y^{\prime}+\frac{i g_v  V_y}{\sqrt{f}}
	\end{pmatrix},	~~~\mathbb{P}_2=2g_{dv} q e^{\chi/2}  A_t V_y  \frac{z h}{f} \sigma_1 ~.~~~~~
	\end{eqnarray}
	 There are four independent solutions since $\xi^{(S)}$ and $\xi^{(C)}$ are four compoments spinor. The general solution can be written in linear combination of all four solutions as 
	\begin{eqnarray}
		\xi^{(S)}= \sum_{i=1}^{4} c_i \xi^{(S,i)} = \mathbb{S}(z)  \mathbf{c}, ~~~~\xi^{(C)}= \sum_{i=1}^{4} c_i \xi^{(C,i)} = \mathbb{C}(z) \mathbf{c}
		\label{eq221}
	\end{eqnarray}
where $\xi^{(S,i)}$ and $\xi^{(C,i)}$ are the $i$-th component of the solution of $\xi^{(S)}$ and $\xi^{(C)}$ respectively. $c_i$ is the $i$-th coefficient for the linear combination of all four solutions. The $4\times4$-matrix $\mathbb{S}(z), \mathbb{C}(z)$ are constructed from the solution, whose $i$-th column is given by $\xi^{(S,i)}$ and $\xi^{(C,i)}$ respectively. The constant column vector $\mathbf{c}$ is constructed from the four coefficients of the linear combination, which is same for both solutions $\xi^{(S)}$ and $\xi^{(C)}$ since we can express $\xi^{(C)} $ in terms of $\xi^{(S)}$ from eq.(s)(\ref{eq215},\ref{eq216}). Substituting above eq.(\ref{eq221}) in eq.(s)(\ref{eq215}, \ref{eq216}), we obtain
	\begin{eqnarray}
		\label{eq223}
		\partial_z \mathbb{S}(z) + \mathbb{M}_1 \mathbb{S}(z)  + \mathbb{M}_2 \mathbb{C}(z) &=& 0 ~,\\
		\partial_z \mathbb{C}(z) + \mathbb{M}_3 \mathbb{C}(z)  + \mathbb{M}_4 \mathbb{S}(z) &=& 0 ~.
		\label{eq224}
	\end{eqnarray}
	From the boundary solutions of the fermion (eq.(\ref{eq211a})),  the asymptotic behaviour of the source and condensation read
	\begin{eqnarray}
		\xi^{(S)}= \begin{pmatrix}
			\mathbf{A} z^m \\
			\tilde{\mathbf{D}}^{*} z^{-m}
		\end{pmatrix}, ~~~~~	\xi^{(C)}= \begin{pmatrix}
			\mathbf{D} z^{-m} \\
			\tilde{\mathbf{A}}^{*} z^{m}
		\end{pmatrix} ~.
		\label{eq224a}
	\end{eqnarray}
	The boundary solution tells us that we need to define $U(z)= diag(z^m, z^m, z^{-m}, z^{-m})$ to get normalized boundary Green's function. Then we can write the boundary solution from eq.(\ref{eq221})
	\begin{eqnarray}
		\xi^{(S)}=  \mathbb{S}(z)  \mathbf{c} \overset{z \rightarrow 0}{\approx } U(z) \mathbb{S}_0 \mathbf{c}, ~~~~\xi^{(C)}=  \mathbb{C}(z) \mathbf{c} \overset{z \rightarrow 0}{\approx } U(z)^{-1} \mathbb{C}_0 \mathbf{c} ~,
		\label{eq225}
	\end{eqnarray}
	where $\mathbb{S}_0, \mathbb{C}_0$ are the $z$-independent boundary $4\times4$-matrix. We can define  
	\begin{eqnarray}
		\mathcal{J} = \mathbb{S}_0 \mathbf{c}, ~~~~~\mathcal{C} = \mathbb{C}_0 \mathbf{c}
		\label{eq228}
	\end{eqnarray}
	which translate the boundary solution (eq.(\ref{eq225})) as
	\begin{eqnarray}
		\xi^{(S)}  \overset{z \rightarrow 0}{\approx } U(z) \mathcal{J}, ~~~~\xi^{(C)} \overset{z \rightarrow 0}{\approx } U(z)^{-1} \mathcal{C} ~.
		\label{eq227}
	\end{eqnarray}
	Comparing eq.(\ref{eq227}) with eq.(\ref{eq224a}), we find 
	\begin{eqnarray}
		\mathcal{J} = \begin{pmatrix}
			\mathbf{A} \\
			\tilde{\mathbf{D}}^{*}
		\end{pmatrix} , ~~~~	\mathcal{C} = \begin{pmatrix}
			\mathbf{D} \\
			\tilde{\mathbf{A}}^{*}
		\end{pmatrix} ~.
	\end{eqnarray}
	From the boundary action (eq.(\ref{eq218})), we can write
	\begin{eqnarray}
		S_{bdy} = \int d^3x \xi^{(S)\dagger} \tilde{\Gamma} \xi^{(C)} + h.c. = \int d^3x \mathcal{J}^{\dagger}\tilde{\Gamma}\mathcal{C} + h.c.
		\label{eq230}
	\end{eqnarray}
	We can also get the relation between $\mathcal{C}$ and $\mathcal{J}$ from eq.(\ref{eq228}) 
	\begin{eqnarray}
		\mathcal{C} = \mathbb{C}_0 \mathbb{S}^{-1}_0 \mathcal{J} ~.
	\end{eqnarray}
	The boundary action now becomes 
	\begin{eqnarray}
		S_{bdy} =  \int d^3x \mathcal{J}^{\dagger}\tilde{\Gamma} \mathbb{C}_0 \mathbb{S}^{-1}_0 \mathcal{J}  + h.c. = \int d^3x \mathcal{J}^{\dagger} \mathbb{G}_0 \mathcal{J} + h.c.
	\end{eqnarray}
	where the retarded Green's function $\mathbb{G}_0= \tilde{\Gamma} \mathbb{C}_0 \mathbb{S}^{-1}_0$. We can promote this boundary Green's function into bulk Green's function by considering the $z$-dependent Green's function as follows:
	\begin{eqnarray}
		\mathbb{G} = \tilde{\Gamma} \mathbb{C}(z) \mathbb{S}^{-1} (z)
	\end{eqnarray}
	where $\mathbb{C}(z), \mathbb{S} (z)$ is defined in eq.(\ref{eq221}). Taking derivative of the above equation, we get
	\begin{eqnarray}
		\partial_z \mathbb{G}(z) =\tilde{\Gamma} \left[\partial_z\mathbb{C}(z)\mathbb{S}^{-1}(z) - \mathbb{C}(z)\mathbb{S}^{-1}(z)(\partial_z\mathbb{S}(z)) \mathbb{S}^{-1}(z) \right] ~.
	\end{eqnarray}
	Using eq.(s)(\ref{eq223},\ref{eq224}), we have found
	\begin{eqnarray}
		\partial_z \mathbb{G}(z) + \tilde{\Gamma} \mathbb{M}_3 \tilde{\Gamma} \mathbb{G}(z) - \mathbb{G}(z) \mathbb{M}_1 -\mathbb{G}(z) \mathbb{M}_2\tilde{\Gamma} \mathbb{G}(z)+ \tilde{\Gamma} \mathbb{M}_4 =0 ~.
		\label{equflowm}
	\end{eqnarray}
	This is the desired flow equation to know the bulk Green's function $\mathbb{G}(z)$. From eq.(\ref{eq225}), we can express 
	\begin{eqnarray}
		\mathbb{S}(z) \overset{z \rightarrow 0}{\approx } U(z) \mathbb{S}_0 ~~\text{and}~~ \mathbb{C}(z) \overset{z \rightarrow 0}{\approx } U(z)^{-1} \mathbb{C}_0 ~.
	\end{eqnarray} 
	By substituting above relations, we can now map the boundary Green's function with bulk Green's function near the boundary in following way
	\begin{eqnarray}
		\mathbb{G} (z) \overset{z \rightarrow 0}{\approx } U(z)^{-1} \mathbb{G}_0 U(z)^{-1}
	\end{eqnarray}
	where we have used the fact $\tilde{\Gamma} U(z)^{-1}\tilde{\Gamma}=U(z)^{-1}  $. To solve the flow equation, we need to know the horizon value of the Green's function which is $\mathbb{G} (z_h) = i \bold{1}_{4\times 4}$. The derivation of $\mathbb{G} (z_h)$ is given in Appendix A. For the numerical evaluation of the Green's function, we will fix the mass of the fermion to be zero ($m_f = 0$).

    \section{Fermionic gap from spectral density}
\begin{figure}[h!]
		\centering
		\begin{subfigure}[b]{0.45\textwidth}
			\centering
			\includegraphics[scale=0.27]{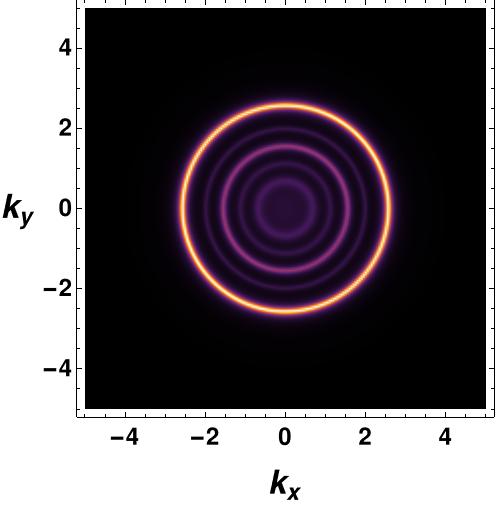}
			\caption{Probe limit case}
		\end{subfigure}
		\hfil
		\begin{subfigure}[b]{0.45\textwidth}
			\centering
			\includegraphics[scale=0.27]{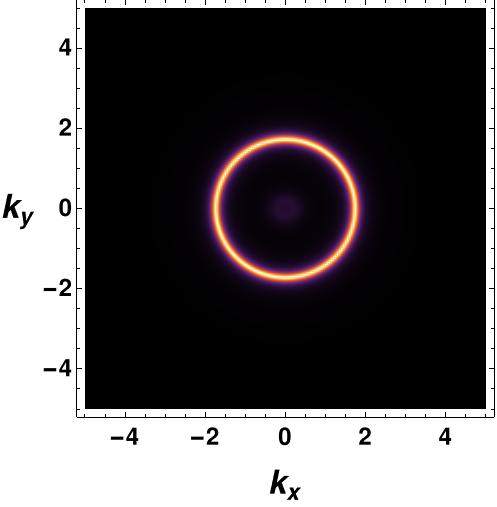}
			\caption{Backreaction case}
		\end{subfigure}
		\caption{Fermionic spectral function with gauge field interaction (only) at $T=0.017\mu$}
		\label{FigSF1}
	\end{figure}

\noindent We now turn our attention to the fermionic spectral function $A(\omega, k_x, k_y)= Tr [Im ~\mathbb{G}(0)]$ in the presence of fully backreacted bosonic fields. Before considering the interactions $\mathcal{L}_{int}$ in (\ref{fermieq1}) with the vector field $V_{\mu}$, we first examine the fermionic spectral function in the presence of only the gauge field $A_{\mu}$, both in the probe limit and the backreacted case.
By numerically solving the flow equation (\ref{equflowm}) using the horizon value of the bulk Green's function (\ref{eqgreenha}), we can obtain the fermionic spectral function. In the probe limit, we consider the AdS-Schwarzschild background, where the backreaction of the bosonic matter field on the background spacetime is neglected.
	The spectral function for $T=0.017\mu$ is depicted in Figure \ref{FigSF1}, which clearly shows that the backreaction contributes to erasing the inside circle of the Fermi surface.  {The possible reason for cleaning the spectral function inside the Fermi surface is as follows. The degree of freedom inside the spectral cone is due to the interaction. 
In the absence of the interaction, all degrees of freedom should sit at the singularity of the Green function, which is the delta function sitting at the dispersion relation. Usually, the fuzzier is the spectrum for the stronger interaction. The fact that backreacted metric gives less fuzzy spectral function means that errors in the solutions seem to be acting as an extra interaction.} 
The radius of the Fermi surface is determined by the chemical potential. An interesting observation is that the radius of the Fermi surface is smaller in the backreaction case. Since there are no interactions besides gauge, the spectral function does not exhibit any gap feature in the $\omega$ vs $k_x (\text{or}~k_y)$ plot. The gap feature only arises when we introduce scalar, vector, or tensor field interactions with the fermion field.
	These results are consistent with previous findings in \cite{Yuk:2022lof}. 
	\subsection{The vector interaction}
\begin{figure}[h!]
		\centering
		\includegraphics[scale=0.25]{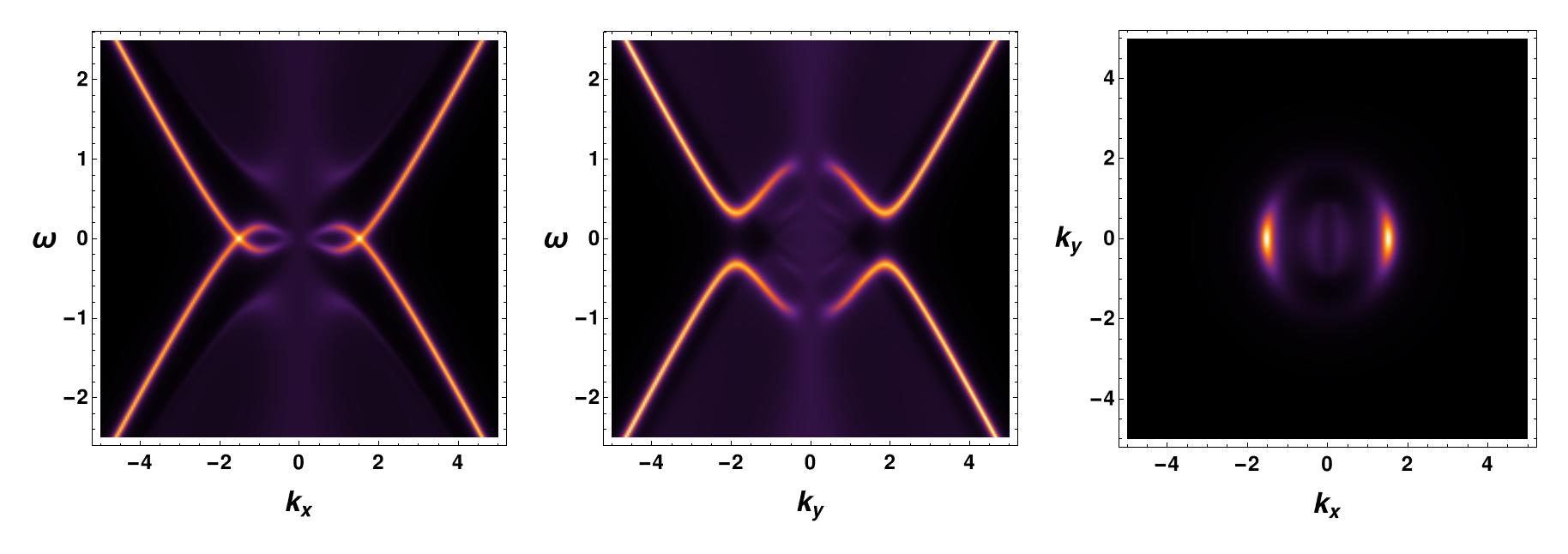}
		\caption{Fermionic spectral function for $g_v=1 $ at $ T=0.017\mu$}
		\label{FigSF2}
	\end{figure}

\begin{figure}[h!]
    	\centering
        \begin{subfigure}[b]{0.45\textwidth}
    	\centering
    	\includegraphics[scale=0.3]{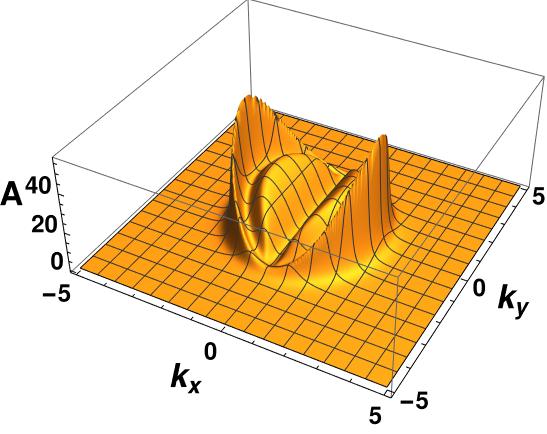}
    	\caption{Probe limit}
    \end{subfigure}
\hfil
\begin{subfigure}[b]{0.45\textwidth}
	\centering
	\includegraphics[scale=0.3]{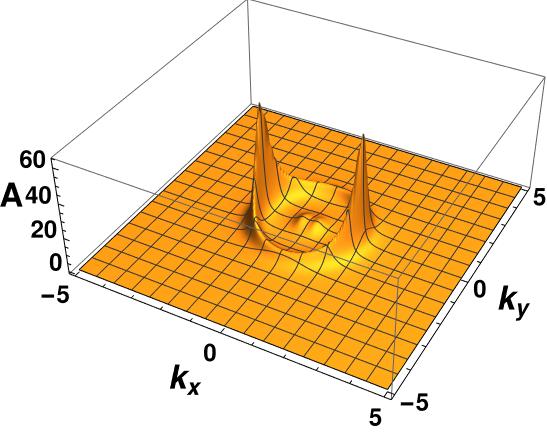}
	\caption{Backreaction}
\end{subfigure}
    	\caption{Comparing spectral function in probe limit and backreaction case.}
    	\label{FigSF3}
    \end{figure}

\begin{figure}[h!]
   	\centering
   	\begin{subfigure}[b]{0.2\textwidth}
   		\centering
   		\includegraphics[scale=0.25]{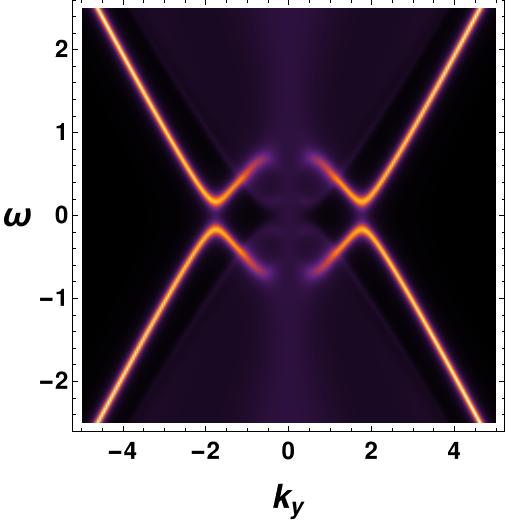}
   		\caption{For $g_v=\frac{1}{2}$}
   	\end{subfigure}
   	\hfil
   	   	\begin{subfigure}[b]{0.2\textwidth}
   		\centering
   		 \includegraphics[scale=0.25]{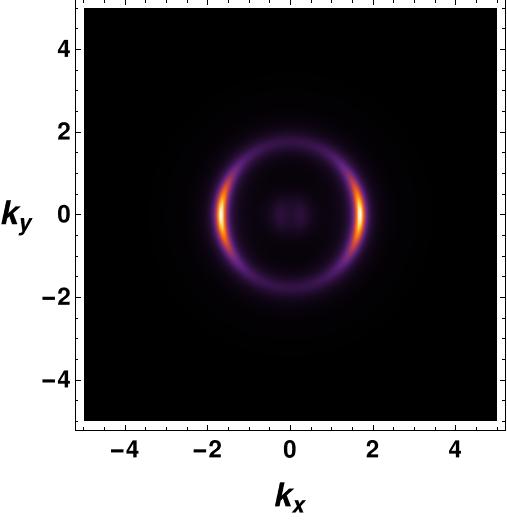}
   		\caption{For $g_v=\frac{1}{2}$}
   	\end{subfigure}
   	\hfil
   	\begin{subfigure}[b]{0.2\textwidth}
   		\centering
   		\includegraphics[scale=0.245]{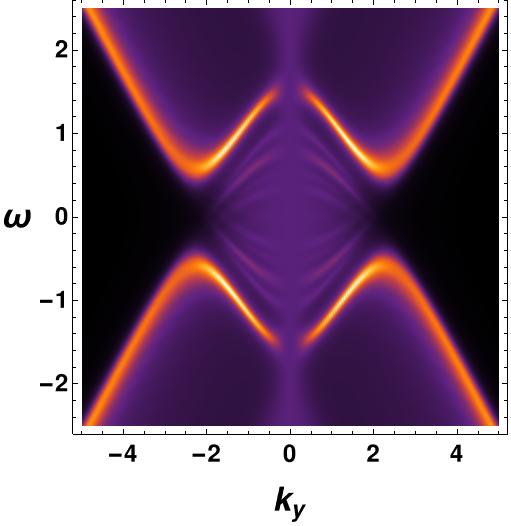}
     \caption{For $g_v=2$}
   	\end{subfigure}
   \hfil
   	\begin{subfigure}[b]{0.2\textwidth}
   		\centering
   		\includegraphics[scale=0.25]{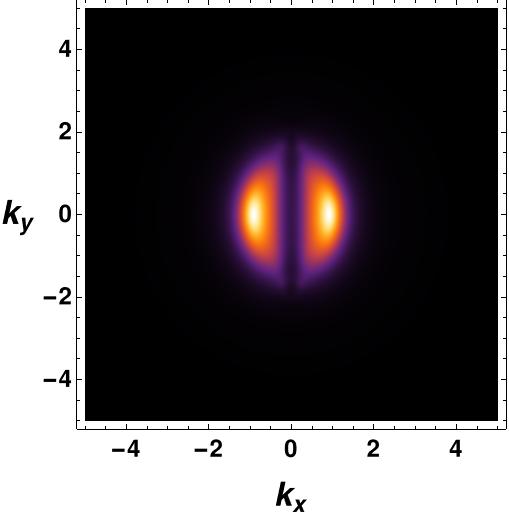}
   		\caption{For $g_v=2$}
   	\end{subfigure}
   	\caption{Dependence of the coupling constant $g_v$ values for fixed $T=0.017\mu$.}
   	\label{FigSF4}
   \end{figure}

     We start by considering only the vector interaction $\bar{\psi} V_{\mu}\Gamma^{\mu} \psi_c$. Setting $\delta_{+}=2$ and $T=0.017\mu$, we plot the fermionic spectral function for $g_v=1$ in Figure \ref{FigSF2}.
     In the $\omega$ vs $k_x$ plot, there is no gap, and a non-vanishing fermionic gap is in the $\omega$ vs $k_y$ plot since the order parameter is zero and maximum at $0^{\circ}$ and $90^{\circ}$ angle in momentum space respectively. Therefore, the $k_x$ vs $k_y${\footnote{All the subsequent $k_x$ vs $k_y$ figures are at $\omega=0$.}} plot at $\omega=0$ shows a Fermi arc along the $k_x$-direction and a fermionic gap along the $k_y$-direction. This is the $p$-wave fermionic gap in the presence of the vector condensate. 
\noindent We now present a comparison of the fermionic spectral function in the probe limit case and the backreaction case, as shown in Figure \ref{FigSF3}, for $g_v=0.5$ at $T=0.017\mu$. The $3d$ figures make it evident that the backreaction has a significant impact on the fermionic spectral function since backreaction helps to erase the non-zero value of the spectral function $A$ inside the arc region. Although there is a slight residue of $A$ within the Fermi arc region even in the backreaction case, it can be disregarded since its value is negligible in comparison to the spectral value at the arc surface. As always, we obtain a gap at $90^{\circ}$ and $270^{\circ}$ angles in momentum space, with a Fermi arc along the $k_x$-direction.
    The effect of the strength of the vector interaction on the fermionic gap is shown in Figure \ref{FigSF4}. As we can see, the gap increases with increasing coupling strength. However, at the same time, the radius of the arc decreases for higher values of coupling constant $g_v$.
   
\begin{figure}[h!]
   	\centering
   	   	\begin{subfigure}[b]{0.3\textwidth}
   		\centering
   		\includegraphics[scale=0.25]{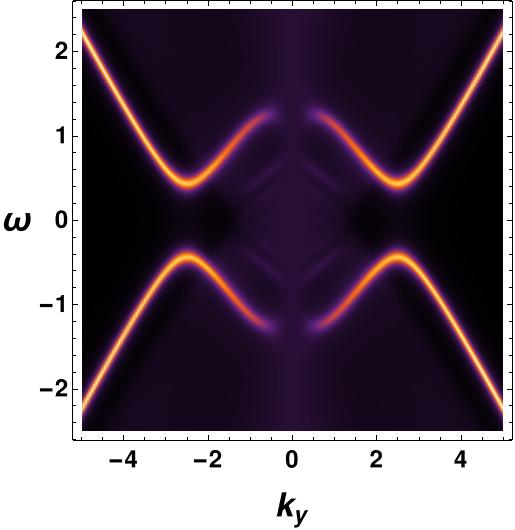}
   		\caption{For $T=0.334 T_c$}
   	\end{subfigure}
   	\hfil
   	\begin{subfigure}[b]{0.3\textwidth}
   		\centering
   		\includegraphics[scale=0.25]{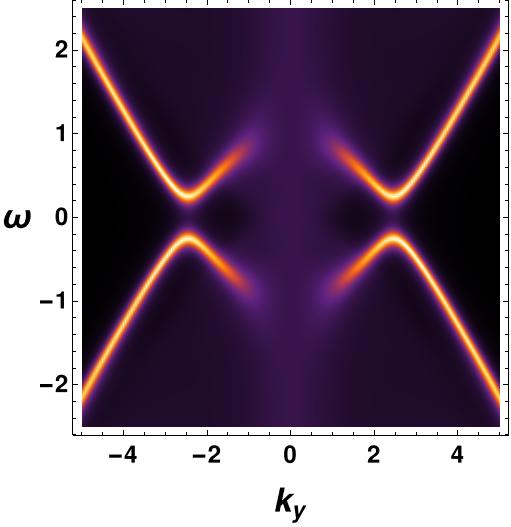}
   		\caption{For $T=0.941 T_c$}
   	\end{subfigure}
   	\hfil
   	\begin{subfigure}[b]{0.3\textwidth}
   		\centering
   		\includegraphics[scale=0.25]{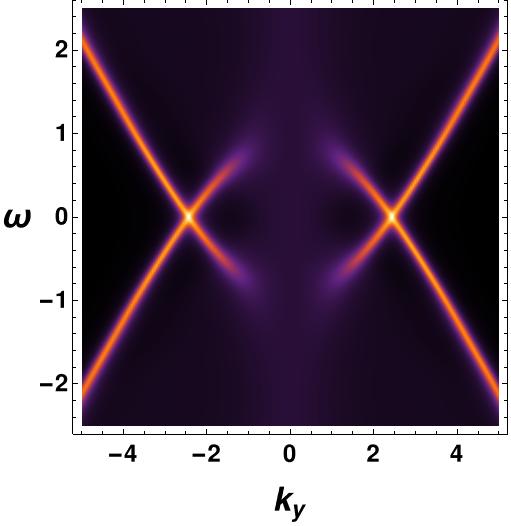}
   		\caption{For $T\geq T_c$}
   	\end{subfigure}
   	\caption{Effect of temperature on fermionic gap for $g_v=1$.}
   	\label{FigSFnw}
   \end{figure}

\noindent We are also interested in understanding the effect of the ratio of temperature to chemical potential on the fermionic spectral function. To investigate this further, we have conducted numerical analyses to examine the $\omega$-gap present in the fermionic spectral function, which is directly proportional to the order parameter value.
   Through the study of bosonic configurations, it has been established that the value of the order parameter exhibits a decline with the rise in temperature, eventually reaching zero at the critical temperature $T_c$ and beyond. Likewise, the fermionic gap experiences a reduction with increasing temperature, reaching zero at $T_c$, shown in Figure \ref{FigSFnw}. As anticipated, the radius demonstrates an upward trend with the augmentation of the chemical potential.
	
	\subsection{The derivative vector interaction}
\begin{figure}[h!]
		\centering
		\includegraphics[scale=0.25]{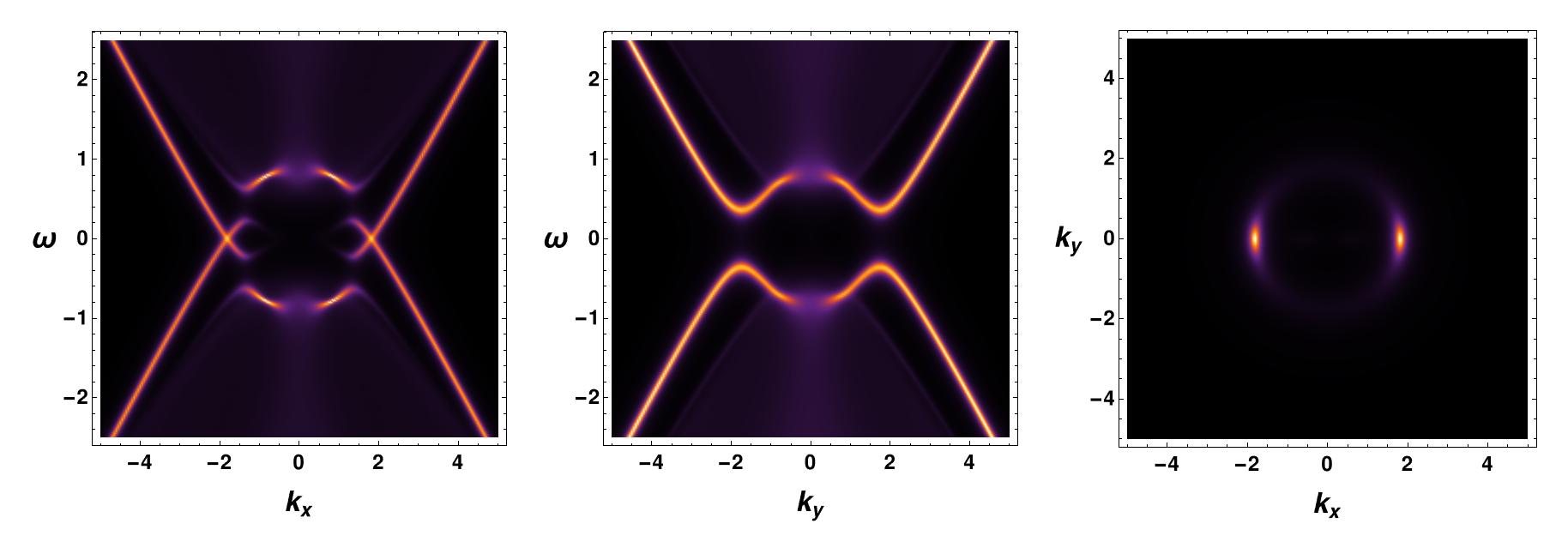}
		\caption{Fermionic spectral function for $g_{dv}=\frac{1}{2}, \delta_{+}=2, T=0.017\mu$}
		\label{FigSF6}
	\end{figure}

\begin{figure}[h]
      	\centering
    	\includegraphics[scale=0.6]{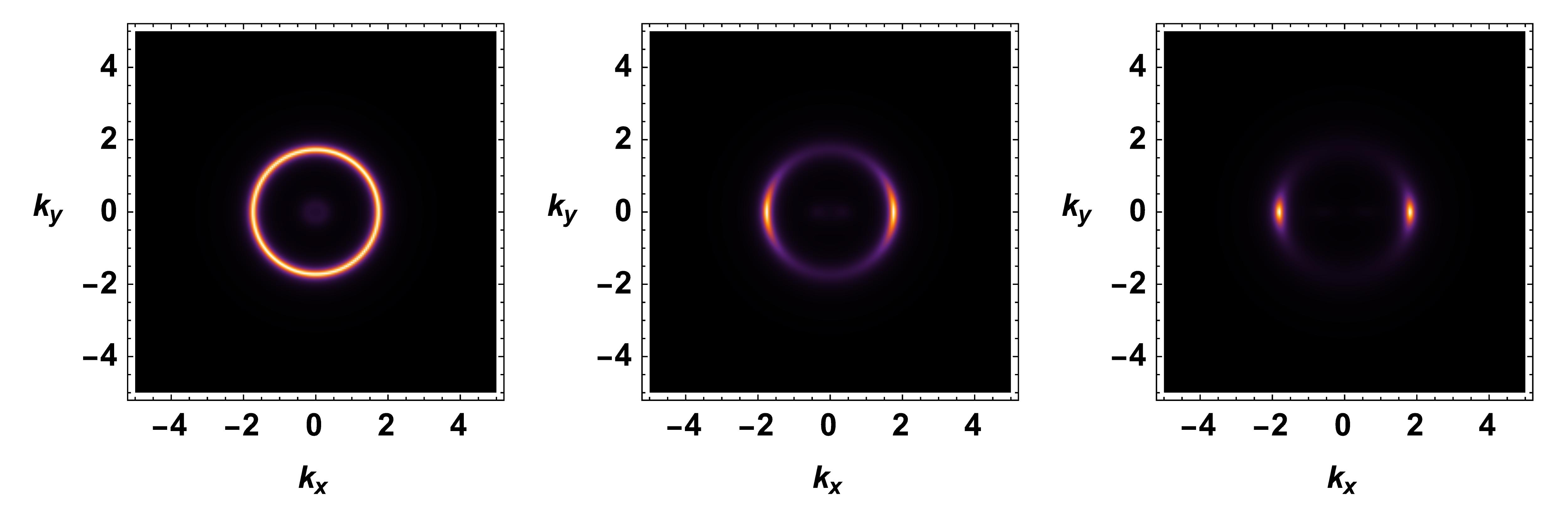}
    \caption{Effect of $g_{dv}=0, \frac{1}{4},\frac{1}{2}$ (left to right) for fixed $T=0.017\mu$.}
    \label{FigSF7}
\end{figure}

\begin{figure}[h]
	\centering
	\includegraphics[scale=0.72]{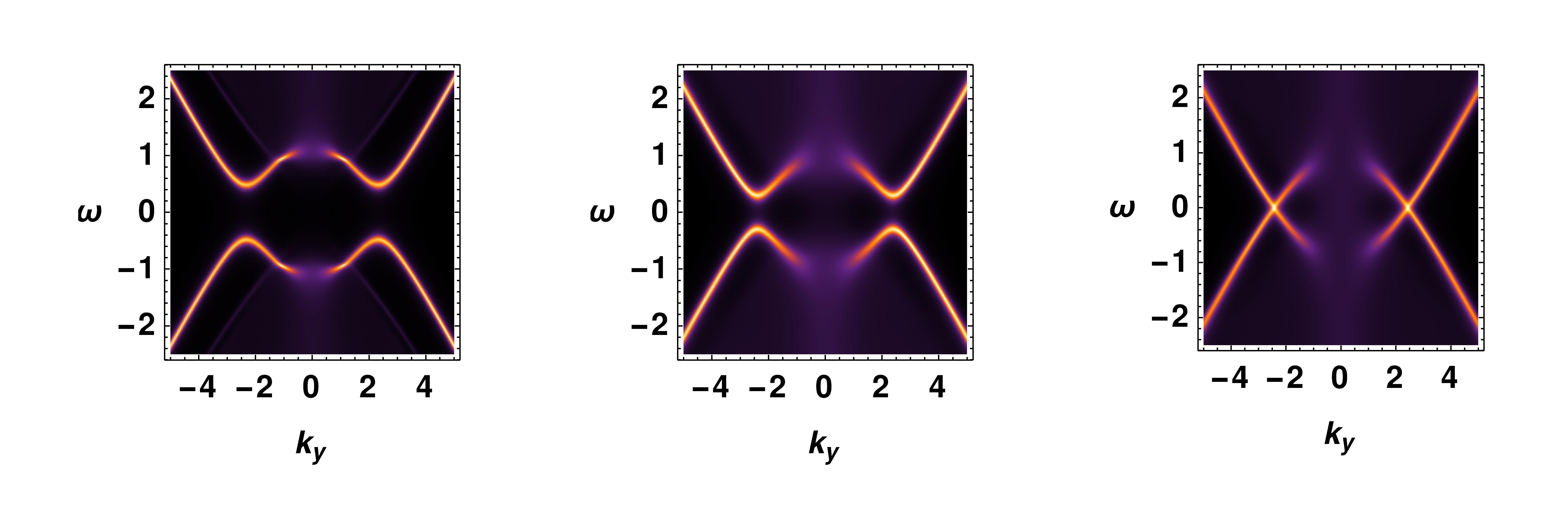}
	\caption{Effect of $T=0.334T_c, 0.941T_c, T_c$ (left to right) for $g_{dv}=\frac{1}{2}$.}
	\label{FigSFT}
\end{figure}

	\noindent We will now investigate the impact of the interaction between the covariant derivative of the vector field and the fermionic field ($g_{dv} \bar{\psi} V_{\mu\nu}\Gamma^{\mu\nu} \psi_c$) on the fermionic spectrum, which also results in a $p$-type fermionic gap. This interaction was studied for a couple of reasons. Firstly, it is known to eliminate the inside region of the spectral function for the vector interaction in the probe limit. Secondly, this interaction yields a more pronounced Fermi arc structure in the spectral function, as demonstrated in Figure \ref{FigSF6}.
    We observe a distinctive high peak in the $\omega$ vs $k_x$ plot along the zero degrees angle in momentum space $(k_y=0)$, which contrasts with the vector interaction. A sharp Fermi arc is visible in this region with a discernible fermionic gap. Notably, the spectral function inside the Fermi arc region is precisely zero, a contrast from the vector interaction, where a very small non-zero spectral function lingers within the Fermi arc region. For higher values of the coupling parameter $g_{dv}$, the fermionic gap increases, and the Fermi arc becomes sharper. Another interesting feature of this coupling is that the radius of the Fermi arc remains constant for any non-zero value of $g_{dv}$ which agrees with the Luttinger sum rule \cite{coleman_2015}, as depicted in Figure \ref{FigSF7}. It appears that the arc shrinks into a minuscule region as the coupling constant $g_{dv}$ value increases.
Since the fermionic gap is proportional to the condensation value, any variations in the temperature to chemical potential ratio ($\frac{T}{\mu}$) have a significant impact on the spectral function, as evidenced by the changes in the fermionic gap. As demonstrated in the spectral function displayed in Figure \ref{FigSFT}, the fermionic gap increases as the temperature declines. Our findings also reveal that, when the temperature is kept constant, an increase in the chemical potential leads to the higher value of the radius of the Fermi arc.

  \subsection{For both interactions}
\begin{figure}[h!]
  	\centering
  	\begin{subfigure}[b]{0.4\textwidth}
  		\centering
  		\includegraphics[scale=0.26]{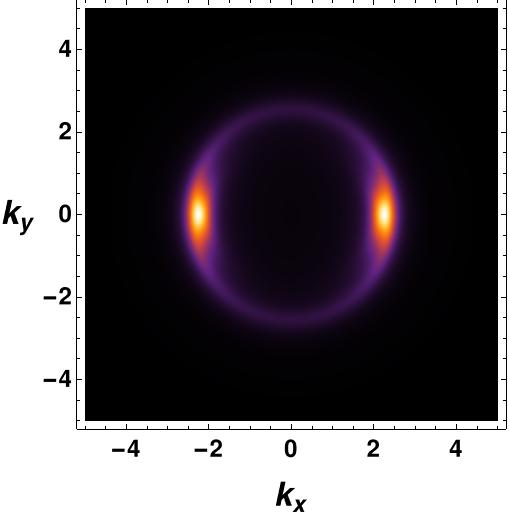}
  		\caption{Probe limit with both interactions}
  	\end{subfigure}
  	\begin{subfigure}[b]{0.4\textwidth}
  		\centering
  		\includegraphics[scale=0.26]{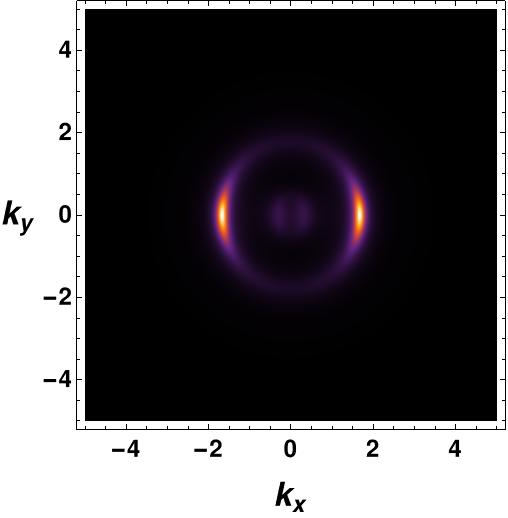}
  		\caption{Backreaction with both interaction}
  	\end{subfigure}
  	\caption{Comparison for fixed $T=0.017\mu, g_v=0.5$ and $g_{dv}=0.1$.}
  	\label{FigSFp2}
  \end{figure}

In this subsection, we examine the combined impact of both interactions on the spectral function, seeking to discern any notable changes. We observe a significant difference in the spectral function when we turn on both interactions. The non-zero value of the spectral function inside the region of the Fermi arc for the vector interaction in the probe limit is gone when we turn on both interactions, shown in Figure \ref{FigSFp2}.
   In the next Figure \ref{FigSF9}, we have compare the spectral function for different combination of the coupling constants values.
\begin{figure}[h!]
  	\centering
  	\begin{subfigure}[b]{0.3\textwidth}
  		\centering
  		\includegraphics[scale=0.26]{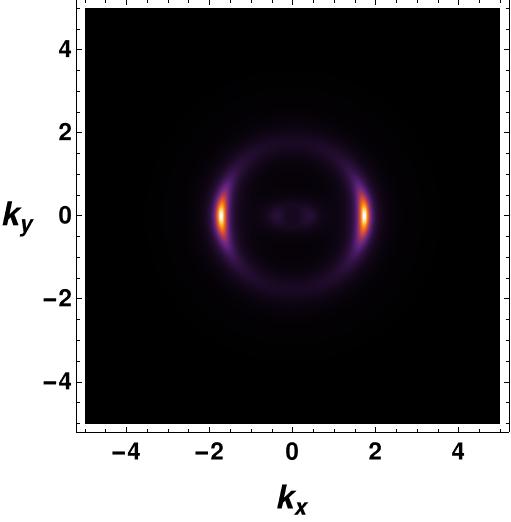}
  		\caption{For $g_v=\frac{1}{4}$ and $g_{dv}=\frac{1}{4}$}
  	\end{subfigure}
  	\hfil
  	\begin{subfigure}[b]{0.3\textwidth}
  		\centering
  		\includegraphics[scale=0.26]{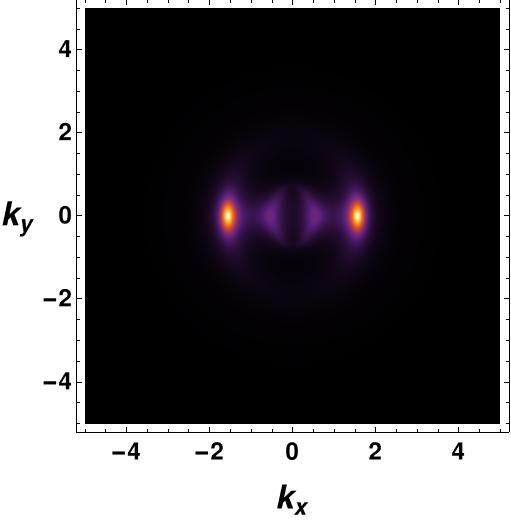}
  		\caption{For $g_v=\frac{3}{4}$ and $g_{dv}=\frac{1}{4}$}
  	\end{subfigure}
  	\hfil
  	\begin{subfigure}[b]{0.3\textwidth}
  		\centering
  		\includegraphics[scale=0.26]{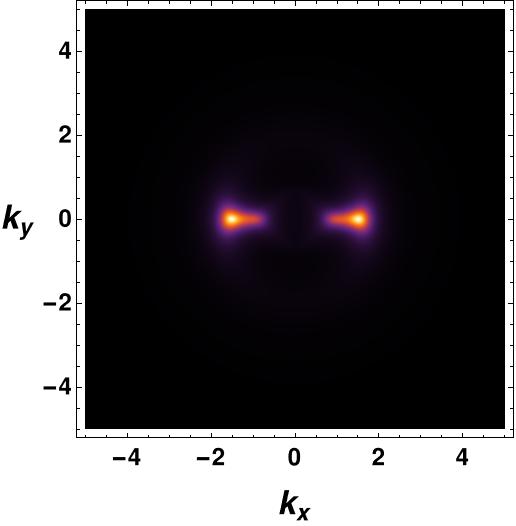}
  		\caption{For $g_v=\frac{3}{4}$ and $g_{dv}=\frac{1}{2}$}
  	\end{subfigure}
  	\caption{Spectral function for fixed $T=0.017\mu$ with different $(g_v, g_{dv})$}
  	\label{FigSF9}
  \end{figure}    
When both interactions are active, we observe that the temperature and chemical potential have a precisely similar influence on the spectral function, just like when each interaction is studied independently. In Figure \ref{FigSF10}, we display the spectral function below and above the critical temperature to illustrate these effects.
 \begin{figure}[h!]
	\centering
	\begin{subfigure}[b]{0.4\textwidth}
		\centering
		\includegraphics[scale=0.26]{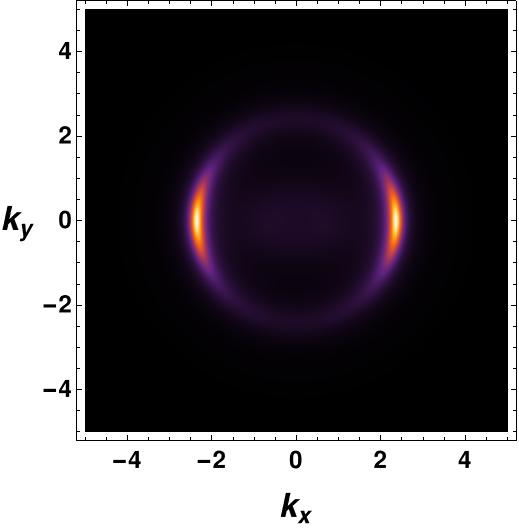}
		\caption{For $T=0.941 T_c$}
	\end{subfigure}
	\begin{subfigure}[b]{0.4\textwidth}
		\centering
		\includegraphics[scale=0.26]{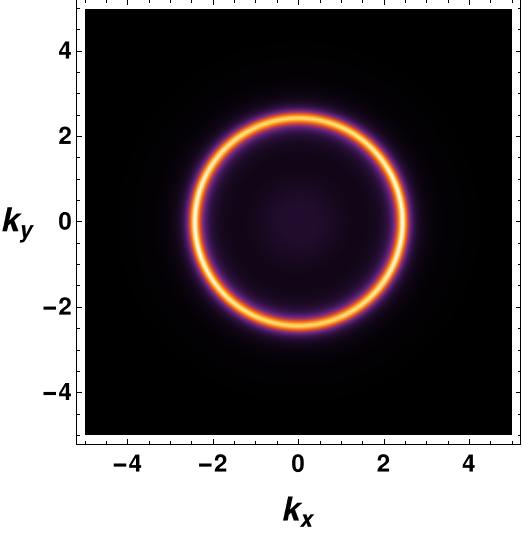}
		\caption{For $T\geq T_c$ }
	\end{subfigure}
	\caption{Comparison below and above $T_c$ for fixed $g_v=0.5$ and $g_{dv}=0.25$.}
	\label{FigSF10}
   \end{figure}

   \section{Summary}
To include the Fermi surface feature in holographic superconductors, we need to study the fermionic spectral function within the holographic setup. In the absence of any bosonic condensate, the holographic setup reveals a Fermi surface whose radius is determined by the chemical potential. 
   In this paper, we have investigated the fermionic spectral function with a full backreaction in the presence of a vector condensate. We determine the critical temperature to be $T_c=0.03748\mu$ for the scaling dimension $\delta_{+}=2$ and obtain all backreacted field configurations below the critical temperature. We have then calculated the ac conductivity in this setup and found $\omega$-gap in the real part of the conductivity.
   Utilizing these fields configurations and two types of interactions, namely vector interaction and the derivative vector interaction, we numerically investigate the fermionic spectral function by solving the flow equation of the bulk Green's function. 
    The motivation for considering these two interactions is that these are the two possible minimal vector field interactions with fermion that produce $p$-wave fermionic spectral function.
   Our analysis explicitly reveals the $p$-wave fermionic gap and Fermi arc in the presence of the vector field.
   The momentum dependent order parameter for $p$-wave holographic superconductors, denoted by $\Delta_k$, exhibits angle dependent features. Upon comparing the momentum dependent order parameter structure (Fig.\ref{figgap}) with the fermionic gap structure (Fig.\ref{FigSF3}(b)), we observe the agreement between them.  {In the momentum-dependent order parameter, we find that its maximum value aligns with the $k_y$ direction and is zero along $k_x$. Similarly, the fermionic spectral function displays a gap along $k_y$ and gapless points (Fermi arc) along $k_x$. Therefore, the spectral function is consistent with the momentum dependent order parameter, confirming the corrected order parameter for holographic superconductors.}
   The $\omega$-gap from the real part of the conductivity and the fermionic spectral function is proportional to the condensation value. These findings could offer valuable insights into the fundamental physics of high-temperature superconductors and provide a foundation for future studies in this field.   \\

\noindent  We also investigate the effect of the coupling constant and the ratio of $\frac{T}{\mu}$ in the fermionic spectral function. 
 The findings of our study indicate that increasing the coupling constant $g_{dv}$ causes the Fermi arc to collapse into a small region along the zero-degree angle in momentum space. However, for higher values of $g_{dv}$, the radius of the Fermi arc remains constant, while it decreases for higher values of the coupling constant $g_{v}$. The shape of the Fermi surface changes with higher values of the vector coupling constant $g_v$. It would be interesting to investigate whether the Luttinger sum rule \cite{coleman_2015} is valid or not in such cases. Notably, the derivative vector interaction yields absolutely zero spectral function values inside the Fermi arc, while the vector interaction results in a non-zero value of the spectral function inside the arc.
   The study also reveals that in the probe limit case, the vector interaction consistently produces a non-zero value of spectral function inside the Fermi surface. However, turning on both interactions causes the value of the spectral function inside the arc to become zero in the probe limit.
   The impact of temperature on the system is an important aspect to consider. It has been observed that as the temperature increases, the condensation value decreases, which in turn leads to a reduction in the gap for both types of interactions. When the system reaches a critical temperature, it undergoes a transition to the normal phase, which is characterized by the appearance of a Fermi surface in the spectral due to the closure of the superconducting gap. All the results in this paper are based on the scaling dimension $\delta_{+}=2$.
   In addition, we have verified the spectral function for the scaling dimension $\delta_{+}=7/4  ~(\text{when}~m^2=-3/16)$, which yields the same conclusions.
   In the future, we will explore $d$-wave holographic superconductors with a fully backreacted metric. This approach could provide valuable insights into the underlying physics of high-temperature superconductors, which exhibit a $d$-wave gap.

  \acknowledgments
  This  work is supported by Mid-career Researcher Program through the National Research Foundation of Korea grant No. NRF-2021R1A2B5B02002603 and NRF-2022H1D3A3A01077468 as well as RS-2023-00218998 of the Basic research Laboratory support program.
  We  thank the APCTP for the hospitality during the focus program, where part of this work was discussed.

	\appendix
	\section{The horizon value of the Green's function}
	\noindent Taking near horizon ansatz of spinor components $\Psi_{i}= \left(1-\frac{z}{z_h}\right)^a \zeta_{i0}$ (where $\zeta_{i0}$ are constants), we can solve Dirac equation (\ref{eq39}) in the near horizon limit which leads two sets of solution of $a$ and the spinor components in following way \cite{Yuk:2022lof}:
	\begin{eqnarray}
		a=\pm \frac{i\omega z_h}{3}, ~~~~ \zeta_{30}=\pm \zeta_{20}, ~~~~ \zeta_{40}=\mp \zeta_{10} ~.
	\end{eqnarray}
	Therefore, the near horizon solution of the spionor reads
	\begin{eqnarray}
		\Psi(z) = \left\{ 
		\begin{array}{ c l }
			\left(1-\frac{z}{z_h}\right)^{-\frac{i\omega z_h}{3}} (\zeta_{10}, \zeta_{20},-\zeta_{20}, \zeta_{10})^T & \quad \textrm{for the infalling ,} \\
			\left(1-\frac{z}{z_h}\right)^{\frac{i\omega z_h}{3}}(\zeta_{10}, \zeta_{20},\zeta_{20}, -\zeta_{10})^T   & \quad \textrm{for the outgoing .} 
		\end{array}
		\right. ~~~~~
	\end{eqnarray}
	Similary, we can get the horizon solution for conjugate fermion. If we consider $\Psi$ has the infalling condition, $\Psi_c$ automatically have the outgoing condition. From the horizon solution of $\Psi$ and $\Psi_c$, we can construct the horizon solution of $\xi^{(S)}(z)$ and $\xi^{(C)}(z)$ using eq.(\ref{eq319}). Then we can construct the $\mathbb{S}(z)$ and $\mathbb{C}(z)$ using eq.(\ref{eq221}) and the horizon solution of $\xi^{(S)}(z)$ and $\xi^{(C)}(z)$. From the definition of the Green's function, we finally obtain the horizon value of the Green's function
	\begin{eqnarray}
		\mathbb{G} (z_h) = \tilde{\Gamma} \mathbb{C} \mathbb{S}^{-1} = i \bold{1}_{4\times 4} ~.
		\label{eqgreenha}
	\end{eqnarray}

\bibliographystyle{jhep}
\bibliography{refpapers.bib}

\end{document}